\documentclass[fleqn,usenatbib]{mnras}
\usepackage{longtable}

\usepackage{newtxtext,newtxmath}

\usepackage[T1]{fontenc}

\DeclareRobustCommand{\VAN}[3]{#2}
\let\VANthebibliography\thebibliography
\def\thebibliography{\DeclareRobustCommand{\VAN}[3]{##3}\VANthebibliography}

\usepackage{graphicx}	
\usepackage{amsmath}	

\title[NW Aps with a potential compact companion star]{An extremely low mass ratio contact binary NW Aps with a potential compact companion star}

\author[Zhou et al.]{
Xiao Zhou,$^{1,2,3}$\thanks{E-mail: zhouxiaophy@ynao.ac.cn}
\\
$^{1}$Yunnan Observatories, Chinese Academy of Sciences (CAS), P.O. Box 110, 650216 Kunming, China\\
$^{2}$Key Laboratory of the Structure and Evolution of Celestial Objects, Chinese Academy of Sciences, P. O. Box 110, 650216 Kunming, China\\
$^{3}$Center for Astronomical Mega-Science, Chinese Academy of Sciences, 20A Datun Road, Chaoyang District, Beijing, 100012, China\\
}

\date{Accepted XXX. Received YYY; in original form ZZZ}

\pubyear{2015}

\begin{document}
\onecolumn
\label{firstpage}
\pagerange{\pageref{firstpage}--\pageref{lastpage}}
\maketitle

\begin{abstract}
The photometric solutions of NW Aps reveal that it is a low mass ratio ($q = 0.086$) contact binary system. Investigation of orbital period shows that its orbital period is increasing continuously at a rate of $dP/dt=+1.117(\pm0.005)\times{10^{-6}}day\cdot year^{-1}$, which may be caused by mass transfer from the less massive component to the more massive one at a rate of $\frac{dM_{2}}{dt}=-3.36(\pm0.02)\times{10^{-8}}M_\odot/year$. A cyclic variation of $P_3 = 22.9(\pm0.1)$ is also found in the O - C curve. There may be a potential compact object orbiting around NW Aps, with its minimum mass to be $M_3 = 0.436(\pm0.007)M_\odot$. However, the magnetic activity of the primary star may also account for the cyclic change. NW Aps is a stellar merger candidate with the longest orbital period among all stellar merge candidates with mass ratio $q < 0.1$. It is still in a stable state since the ratio of orbital angular momentum ($J_{orb}$) to spin angular momentum ($J_{spin}$) is $\frac{J_{orb}}{J_{spin}}$ = 3.257. Both of its primary and secondary star are oversized to main sequence stars, and the surface gravity of the primary and secondary stars are significant lower than main sequence stars. The P - log g relationship is fitted with parabola for low mass ratio contact binary systems. More targets laid in the gap are needed to confirm the P - log g relationship and reveal the final evolutionary state of low mass ratio contact binary system. 
\end{abstract}




\section{Introduction}
\label{sect:intro}
Stellar mergers are predicted to be common events in the Galaxy \citep{2014MNRAS.443.1319K}. However, only a handful number of these transients have been noted so far, which are the merger events of V4332 Sgr \citep{1994IAUC.5942....1H}, V838 Mon \citep{2002IAUC.7785....1B}, V1309 Sco \citep{2008CBET.1496....1N}, and OGLE-2002-BLG-360 \citep{2013A&A...555A..16T}. Common envelope evolution and stellar merger is a very important formation channel for peculiar stars, following with luminous transients \citep{2013A&ARv..21...59I,2016MNRAS.455.4351P}. For example, the formation of blue stragglers have been debated for a long time \citep{2008MNRAS.387.1416C}, and one possible explanation is that they are products of binary merger. The merger event of V1309 Sco was observed on 2008. Its light curves in I band were observed by OGLE on 2004, which revealed that it was a contact binary with extremely low mass ratio (q = 0.094) before merger \citep{2016RAA....16...68Z}. V1309 Sco was observed with infrared CCD recently, which showed that reasonable equilibrium in this stellar merger remnant was being reached in about 9 years after the outburst. V1309 Sco has settled into a nearly constant magnitude, resembling a normal blue star. The asymptotic blue color of V1309 Sco as the resultant of a stellar merger event suggests that the object is a blue straggler in the making, as theoretically predicted \citep{2019MNRAS.486.1220F}.

The discovery of the luminous red nova V1309 Sco (Nova Scorpii 2008) and the fact that its progenitor was an extreme mass ratio contact binary system inspired researchers to search stellar merger candidates in low mass ratio overcontact binary system \citep{2011AJ....141..151Q,2021MNRAS.502.2879G,2022MNRAS.512.1244C}. Owe to the large sky survey projects such as the Optical Gravitational Lensing Experiment (OGLE) \citep{1992AcA....42..253U}, All Sky Automated Survey (ASAS) \citep{1997AcA....47..467P}, Wide-field Infrared Survey Explorer (WISE) \citep{2018ApJS..237...28C}, Transiting Exoplanet Survey Satellite (TESS) \citep{2015JATIS...1a4003R,2022ApJS..258...16P} and Zwicky Transient Facility (ZTF) \citep{2019PASP..131a8002B,2020ApJS..249...18C}, the light curves of contact binaries are accumulating rapidly. However, only a few tens of low mass ratio contact binaries have been identified \citep{2015AJ....150...69Y,2022MNRAS.512.1244C}. \citet{1980A&A....92..167H} revealed that low mass ratio contact binary system with a decreasing period will evolve into a rapidly rotating single star when the photosphere surface of the binary system was close to the outer critical Roche lobe, while system with an increasing period may merge when it met the criterion that the orbital angular momentum was less than 3 times of the total spin angular momentum. Theoretical models predicted that a contact binary system will unstable and merge when its mass ratio reached around $q \sim 0.07 - 0.09$ \citep{2007MNRAS.377.1635A}.

In the present work, we will report our research of a low mass ratio contact binary system NW Aps ($V$ = $9^{m}.08 - 9^{m}.37$, $P = 1.065558$  days), which has the longest orbital period in all low mass ratio contact binary systems to date. Its light variability was first noted by \citet{1969IBVS..330....1S}, named as HD 130338. The target was neglected until its V band light curve from ASAS was analysed by \citet{2005Ap&SS.300..329W}. \citet{2005Ap&SS.300..329W} reported that the mass ratio of NW Aps was $q= 0.100(\pm0.002)$, making it a stellar merger candidate. No more detailed study was carried out after that. Fortunately, it is a quite bright star listed in TESS two-minute target lists as TIC 359987981. Thus, we can acquire its full light curves with very high precision. Its light curves from TESS are displayed in Section 2, and orbital period variations are investigate in Section 3. The light curves are modelled in Section 4. The Spectral Energy Distribution of NW Aps is synthesized in Section 5. Discussions and Conclusions are arranged in Section 6 and Section 7, respectively.

\section{TESS Observations}
\label{sect:Obs}

TESS is a sky survey telescope optimized for detecting exoplanets, which was launched on April, 2018. It is equipped with four cameras which covers a quite large field of view (24 $\times$ 96 degree), and use a wide bandpass filter with its wavelength ranging from nearly 600 to 1000 nm. TESS has been monitoring the whole sky for over six years. The observations for each sector lasting for about 27.4 days, which makes it to be a very suitable telescope for searching close binaries.

\begin{figure}
\centering
\includegraphics[width=0.8\linewidth]{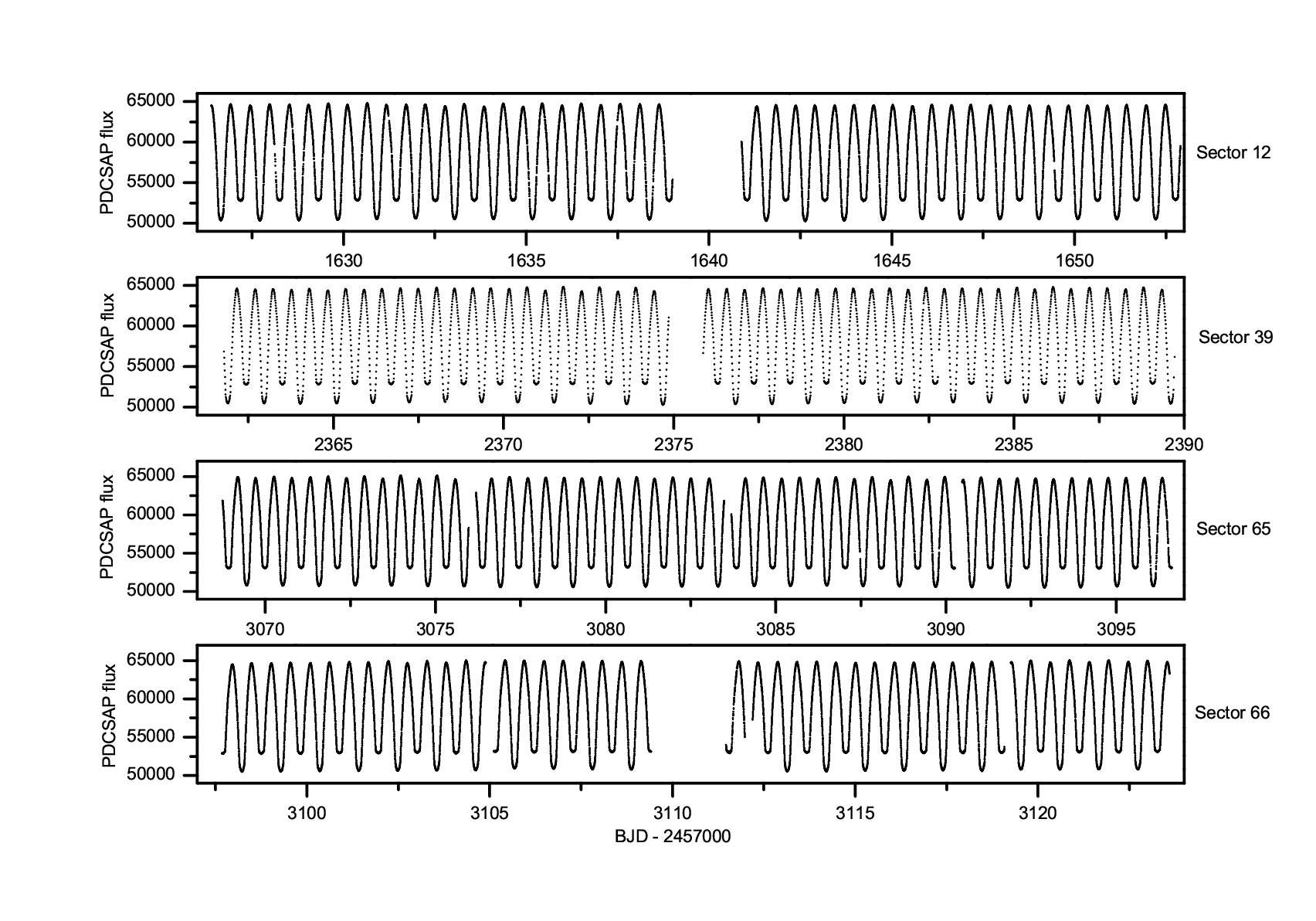}
\caption{Light curves of TIC 359987981 from Sector 12, 39, 65 and 66.}
\label{lc}
\end{figure}

TIC 359987981 was observed by TESS with 2 minutes cadence in Sector 12 extending from 2019 May 21 to 2019 June 18, Sector 65 extending from 2023 May 4 to 2023 June 1, Sector 66 extending from 2023 June 2 to 2023 July 1, and with 10 minutes cadence in Sector 39 extending from May 27 through June 24, 2021. The light curves are downloaded from the MAST data archive (https://archive.stsci.edu/) using the Python package lightkurve \citep{2018ascl.soft12013L}. The light curves extracted with TESS data processing pipeline developed by the Science Processing Operations Center (SPOC) with an exposure time of 120 seconds are downloaded for Sector 12,  Sector 65 and  Sector 66 \citep{2016SPIE.9913E..3EJ}. The light curves extracted with TESS Light Curves From Full Frame Images (TESS-SPOC) pipeline \citep{2020RNAAS...4..201C} with an exposure time of 600 seconds are downloaded for Sector 39. Both the Simple Aperture Photometry (SAP) flux and Pre-search Data Conditioning SAP flux (PDCSAP) flux are provided in TESS light curves. The SAP flux is calculated by summing the brightness of all chosen pixels, while long term trends have been removed in the PDCSAP flux. Thus, PDCSAP flux is usually cleaner than the SAP flux and will have fewer systematic trends. The Pre-search Data Conditioning Simple Aperture Photometry (PDCSAP) flux are used in the present work. The observational light curves of TIC 359987981 are displayed in Fig. \ref{lc}.  

\section{Investigation of orbital period variations}
\label{sect:period}
Contact binaries are typical interaction binary systems. Due to the possible mass transfer between their component stars, long term period changes commonly appear in contact binary systems. \citet{1963ApJ...138..471H} discussed mass ejection in binary systems and predicted their effect on the orbital period of the systems. \citet{2001MNRAS.328..635Q} reported that a relation between the period change and the mass ratio may exist, basing on the research of the period changes of 30 W-type contact binaries. The period of NW Aps was given to be P = 1.065558 days by \citet{2005Ap&SS.300..329W}. Times of minima determined by other researchers are collected, which are HJD 2448500.6490 \citep{2009OEJV..116....1P}, HJD 2451967.9590 (O - C gateway, http://var2.astro.cz/ocgate/), HJD 2452728.7720 (GCVS, http://www.sai.msu.su/gcvs/gcvs/), HJD 2454984.5480 and HJD 2455390.5250 \citep{2010OEJV..130....1P}. A total of 186 times of minima are determined from the light curves displayed in Fig. \ref{lc}, which are all listed in Table \ref{OC}. And the times of light minima are computed using the method introduced by \citet{1956BAN....12..327K}. The Kwee \& van Woerden (K - W) method is a classical and widely used technique for determining the times of light minima in eclipsing binary light curves. Before the computing, we should choose a phase interval from a single eclipse’s observational data, which covers both the descending and ascending branches. Then, we interpolate the observational data to $2n + 1$ equally spaced points with a time interval of $\Delta t$, and an initial epoch of minimum $T_1$ is selected. The number of interpolated in-eclipse observations $2n + 1$ is recommended to be about equal to the total number of observations in the phase interval. Take $T_1$ as reflection axis and reflect the interpolated magnitudes of the descending branch to the ascending branch, and calculate the magnitude differences $\Delta m_k$ ($ k = 1, \ldots, n $) to get $s(T_1) = \sum_{k=1}^n (\Delta m_k)^2$. Assuming $s(T) = a T^2 + b T + c$, and shift the reflection axis to $T_1 + \frac{1}{2} \Delta t$ and $T_1 - \frac{1}{2} \Delta t$ to compute $s(T_1 + \frac{1}{2} \Delta t)$ and $s(T_1 - \frac{1}{2} \Delta t)$. Using the three $s$-values corresponding to $ T_1 - \frac{1}{2} \Delta t $, $ T_1 $ and $ T_1 + \frac{1}{2} \Delta t $, the coefficients $ a, b, c $ will be determined. The Epoch of light minimum is calculated as $T_0 = -\frac{b}{2a}$ and its mean error is given by
$\sigma_{T_0}^2 = \frac{4ac - b^2}{4a^2 (Z - 1)}$, where $ Z $ is the number of independent magnitude pairs. The times of minima determined from the TESS data and those collected from literature are used to investigate the period variations of NW Aps. The minima in HJD are converted to BJD since TESS used BJD time system and all times of minima are listed in Table \ref{OC}. All times of minima are determined from observations obtained with charge-coupled device (CCD) photometry. The time of primary light minimum BJD 2458626.6484 is adopted as the initial time since it is determined with TESS observations, which has very high accuracy. Thus, the linear equation used to calculate the $O - C$ values is:
\begin{equation}
\label{Epoch_O-C}
Min.I(BJD) = 2458626.6484+1^{d}.065558\times{E}.
\end{equation}

The column to column description of Table \ref{OC} are:

Column 1  - Barycentric Julian Date of the observed times of minima (BJD - 2400000);

Column 2  - primary (p) or secondary (s) minima;

Column 3  - error of times of minima;

Column 4  - cycle numbers from the initial epoch (E, as used in Equation \ref{Epoch_O-C});

Column 5  - the $O - C$ values calculated from linear Equation \ref{Epoch_O-C};

Column 6  - the references;

\begin{longtable}{llllllllllll}
\caption{Times of minima and $O-C$ values for NW Aps.}\label{OC}
\setlength{\tabcolsep}{4pt}
\endfirsthead
\multicolumn{12}{l}{Table \ref{OC} continued }\\\hline\hline
    BJD     &   p/s    &  Error    &   Epoch    &     $O - C$    &    Ref. &       BJD     &   p/s    &  Error    &   Epoch    &     $O - C$    &    Ref.  \\
(2400000+)  &          &  (days)   &   (days)   &      (days)    &         &   (2400000+)  &          &  (days)   &   (days)   &      (days)    &          \\\hline
\endhead
\hline\hline
\endfoot
\endlastfoot
\hline\hline
    BJD     &   p/s    &  Error    &   Epoch    &     $O - C$    &    Ref. &       BJD     &   p/s    &  Error    &   Epoch    &     $O - C$    &    Ref.  \\
(2400000+)  &          &  (days)   &   (days)   &      (days)    &         &   (2400000+)  &          &  (days)   &   (days)   &      (days)    &          \\\hline
48500.6497  & p        &           &  -9503     &     -0.0011    &    1    &   59387.4741  &   p      &  0.0007   &    714     &      0.0172    &   5      \\
51967.9598  & p        &           &  -6249     &     -0.0167    &    2    &   59388.0083  &   s      &  0.0007   &    714.5   &      0.0187    &   5      \\
52728.7728  & p        &           &  -5535     &     -0.0122    &    3    &   59388.5396  &   p      &  0.0006   &    715     &      0.0172    &   5      \\
54984.5488  & p        & 0.0100    &  -3418     &     -0.0225    &    4    &   59389.0737  &   s      &  0.0007   &    715.5   &      0.0185    &   5      \\
55390.5258  & p        & 0.0100    &  -3037     &     -0.0230    &    4    &   60068.9193  &   s      &  0.0003   &    1353.5  &      0.0381    &   5      \\
58626.6484  & p        & 0.0003    &  0         &     0          &    5    &   60069.4515  &   p      &  0.0003   &    1354    &      0.0375    &   5      \\
58627.1828  & s        & 0.0003    &  0.5       &     0.0016     &    5    &   60069.9848  &   s      &  0.0003   &    1354.5  &      0.0380    &   5      \\
58627.7144  & p        & 0.0003    &  1         &     0.0004     &    5    &   60070.5168  &   p      &  0.0003   &    1355    &      0.0373    &   5      \\
58628.2480  & s        & 0.0004    &  1.5       &     0.0012     &    5    &   60071.0503  &   s      &  0.0003   &    1355.5  &      0.0380    &   5      \\
58628.7809  & p        & 0.0003    &  2         &     0.0013     &    5    &   60071.5820  &   p      &  0.0003   &    1356    &      0.0369    &   5      \\
58629.3132  & s        & 0.0003    &  2.5       &     0.0009     &    5    &   60072.1166  &   s      &  0.0003   &    1356.5  &      0.0387    &   5      \\
58629.8461  & p        & 0.0003    &  3         &     0.0010     &    5    &   60072.6473  &   p      &  0.0003   &    1357    &      0.0367    &   5      \\
58630.3790  & s        & 0.0003    &  3.5       &     0.0011     &    5    &   60073.1823  &   s      &  0.0003   &    1357.5  &      0.0389    &   5      \\
58630.9119  & p        & 0.0003    &  4         &     0.0012     &    5    &   60073.7122  &   p      &  0.0003   &    1358    &      0.0360    &   5      \\
58631.4448  & s        & 0.0003    &  4.5       &     0.0014     &    5    &   60074.2483  &   s      &  0.0003   &    1358.5  &      0.0393    &   5      \\
58631.9765  & p        & 0.0003    &  5         &     0.0003     &    5    &   60074.7779  &   p      &  0.0003   &    1359    &      0.0362    &   5      \\
58632.5108  & s        & 0.0003    &  5.5       &     0.0018     &    5    &   60075.3140  &   s      &  0.0003   &    1359.5  &      0.0395    &   5      \\
58633.0416  & p        & 0.0003    &  6         &     -0.0002    &    5    &   60075.8438  &   p      &  0.0003   &    1360    &      0.0364    &   5      \\
58633.5767  & s        & 0.0003    &  6.5       &     0.0021     &    5    &   60076.3792  &   s      &  0.0003   &    1360.5  &      0.0391    &   5      \\
58634.1068  & p        & 0.0003    &  7         &     -0.0005    &    5    &   60076.9097  &   p      &  0.0003   &    1361    &      0.0369    &   5      \\
58634.6427  & s        & 0.0003    &  7.5       &     0.0026     &    5    &   60077.4447  &   s      &  0.0003   &    1361.5  &      0.0390    &   5      \\
58635.1721  & p        & 0.0003    &  8         &     -0.0008    &    5    &   60077.9759  &   p      &  0.0003   &    1362    &      0.0374    &   5      \\
58635.7083  & s        & 0.0003    &  8.5       &     0.0026     &    5    &   60078.5101  &   s      &  0.0003   &    1362.5  &      0.0389    &   5      \\
58636.2384  & p        & 0.0003    &  9         &     -0.0001    &    5    &   60079.0416  &   p      &  0.0003   &    1363    &      0.0376    &   5      \\
58636.7735  & s        & 0.0003    &  9.5       &     0.0022     &    5    &   60079.5751  &   s      &  0.0003   &    1363.5  &      0.0384    &   5      \\
58637.3044  & p        & 0.0003    &  10        &     0.0003     &    5    &   60080.1079  &   p      &  0.0003   &    1364    &      0.0383    &   5      \\
58637.8388  & s        & 0.0003    &  10.5      &     0.0020     &    5    &   60080.6405  &   s      &  0.0003   &    1364.5  &      0.0382    &   5      \\
58638.3702  & p        & 0.0003    &  11        &     0.0007     &    5    &   60081.1732  &   p      &  0.0003   &    1365    &      0.0381    &   5      \\
58641.0356  & s        & 0.0003    &  13.5      &     0.0021     &    5    &   60081.7061  &   s      &  0.0003   &    1365.5  &      0.0382    &   5      \\
58641.5664  & p        & 0.0003    &  14        &     0.0001     &    5    &   60082.2387  &   p      &  0.0003   &    1366    &      0.0380    &   5      \\
58642.1008  & s        & 0.0003    &  14.5      &     0.0018     &    5    &   60082.7725  &   s      &  0.0003   &    1366.5  &      0.0391    &   5      \\
58642.6323  & p        & 0.0003    &  15        &     0.0005     &    5    &   60083.3035  &   p      &  0.0003   &    1367    &      0.0373    &   5      \\
58643.1657  & s        & 0.0003    &  15.5      &     0.0011     &    5    &   60083.8382  &   s      &  0.0003   &    1367.5  &      0.0392    &   5      \\
58643.6980  & p        & 0.0003    &  16        &     0.0006     &    5    &   60084.3689  &   p      &  0.0003   &    1368    &      0.0371    &   5      \\
58644.2321  & s        & 0.0003    &  16.5      &     0.0020     &    5    &   60084.9038  &   s      &  0.0003   &    1368.5  &      0.0393    &   5      \\
58644.7640  & p        & 0.0003    &  17        &     0.0011     &    5    &   60085.4342  &   p      &  0.0003   &    1369    &      0.0369    &   5      \\
58645.2984  & s        & 0.0003    &  17.5      &     0.0027     &    5    &   60085.9694  &   s      &  0.0003   &    1369.5  &      0.0393    &   5      \\
58645.8292  & p        & 0.0003    &  18        &     0.0007     &    5    &   60086.4998  &   p      &  0.0003   &    1370    &      0.0369    &   5      \\
58646.3635  & s        & 0.0003    &  18.5      &     0.0022     &    5    &   60087.0350  &   s      &  0.0003   &    1370.5  &      0.0394    &   5      \\
58646.8948  & p        & 0.0003    &  19        &     0.0007     &    5    &   60087.5654  &   p      &  0.0003   &    1371    &      0.0370    &   5      \\
58647.4287  & s        & 0.0003    &  19.5      &     0.0019     &    5    &   60088.1008  &   s      &  0.0003   &    1371.5  &      0.0396    &   5      \\
58647.9605  & p        & 0.0003    &  20        &     0.0009     &    5    &   60088.6310  &   p      &  0.0003   &    1372    &      0.0370    &   5      \\
58648.4939  & s        & 0.0003    &  20.5      &     0.0016     &    5    &   60089.1666  &   s      &  0.0003   &    1372.5  &      0.0398    &   5      \\
58649.0256  & p        & 0.0003    &  21        &     0.0005     &    5    &   60089.6967  &   p      &  0.0003   &    1373    &      0.0371    &   5      \\
58649.5591  & s        & 0.0003    &  21.5      &     0.0011     &    5    &   60090.7625  &   p      &  0.0003   &    1374    &      0.0374    &   5      \\
58650.0916  & p        & 0.0003    &  22        &     0.0009     &    5    &   60091.2972  &   s      &  0.0003   &    1374.5  &      0.0393    &   5      \\
58650.6257  & s        & 0.0003    &  22.5      &     0.0022     &    5    &   60091.8284  &   p      &  0.0003   &    1375    &      0.0377    &   5      \\
58651.1567  & p        & 0.0003    &  23        &     0.0004     &    5    &   60092.3625  &   s      &  0.0003   &    1375.5  &      0.0390    &   5      \\
58651.6911  & s        & 0.0003    &  23.5      &     0.0021     &    5    &   60092.8944  &   p      &  0.0003   &    1376    &      0.0382    &   5      \\
58652.2227  & p        & 0.0003    &  24        &     0.0008     &    5    &   60093.4278  &   s      &  0.0003   &    1376.5  &      0.0387    &   5      \\
58652.7568  & s        & 0.0003    &  24.5      &     0.0022     &    5    &   60093.9601  &   p      &  0.0003   &    1377    &      0.0383    &   5      \\
59362.4342  & s        & 0.0007    &  690.5     &     0.0180     &    5    &   60094.4935  &   s      &  0.0003   &    1377.5  &      0.0389    &   5      \\
59362.9658  & p        & 0.0006    &  691       &     0.0168     &    5    &   60095.0256  &   p      &  0.0003   &    1378    &      0.0382    &   5      \\
59363.4999  & s        & 0.0007    &  691.5     &     0.0181     &    5    &   60095.5597  &   s      &  0.0003   &    1378.5  &      0.0396    &   5      \\
59364.0312  & p        & 0.0006    &  692       &     0.0166     &    5    &   60096.0907  &   p      &  0.0003   &    1379    &      0.0377    &   5      \\
59364.5661  & s        & 0.0007    &  692.5     &     0.0188     &    5    &   60098.2220  &   p      &  0.0003   &    1381    &      0.0379    &   5      \\
59365.0969  & p        & 0.0007    &  693       &     0.0168     &    5    &   60098.7557  &   s      &  0.0003   &    1381.5  &      0.0389    &   5      \\
59365.6307  & s        & 0.0007    &  693.5     &     0.0178     &    5    &   60099.2875  &   p      &  0.0003   &    1382    &      0.0379    &   5      \\
59366.1629  & p        & 0.0007    &  694       &     0.0172     &    5    &   60099.8218  &   s      &  0.0003   &    1382.5  &      0.0394    &   5      \\
59366.6964  & s        & 0.0007    &  694.5     &     0.0180     &    5    &   60100.3533  &   p      &  0.0003   &    1383    &      0.0382    &   5      \\
59367.2282  & p        & 0.0007    &  695       &     0.0169     &    5    &   60100.8872  &   s      &  0.0003   &    1383.5  &      0.0392    &   5      \\
59367.7615  & s        & 0.0007    &  695.5     &     0.0175     &    5    &   60101.4191  &   p      &  0.0003   &    1384    &      0.0384    &   5      \\
59368.2942  & p        & 0.0007    &  696       &     0.0174     &    5    &   60101.9527  &   s      &  0.0003   &    1384.5  &      0.0392    &   5      \\
59368.8275  & s        & 0.0007    &  696.5     &     0.0179     &    5    &   60102.4854  &   p      &  0.0003   &    1385    &      0.0391    &   5      \\
59369.3589  & p        & 0.0007    &  697       &     0.0165     &    5    &   60103.0179  &   s      &  0.0003   &    1385.5  &      0.0389    &   5      \\
59369.8932  & s        & 0.0007    &  697.5     &     0.0180     &    5    &   60103.5507  &   p      &  0.0003   &    1386    &      0.0389    &   5      \\
59370.4237  & p        & 0.0007    &  698       &     0.0158     &    5    &   60104.0837  &   s      &  0.0003   &    1386.5  &      0.0391    &   5      \\
59370.9601  & s        & 0.0007    &  698.5     &     0.0194     &    5    &   60104.6165  &   p      &  0.0003   &    1387    &      0.0391    &   5      \\
59371.4892  & p        & 0.0007    &  699       &     0.0157     &    5    &   60105.6816  &   p      &  0.0003   &    1388    &      0.0386    &   5      \\
59372.0253  & s        & 0.0007    &  699.5     &     0.0190     &    5    &   60106.2153  &   s      &  0.0003   &    1388.5  &      0.0396    &   5      \\
59372.5548  & p        & 0.0007    &  700       &     0.0158     &    5    &   60106.7469  &   p      &  0.0003   &    1389    &      0.0384    &   5      \\
59373.0913  & s        & 0.0007    &  700.5     &     0.0195     &    5    &   60107.2813  &   s      &  0.0003   &    1389.5  &      0.0400    &   5      \\
59373.6207  & p        & 0.0007    &  701       &     0.0161     &    5    &   60107.8119  &   p      &  0.0003   &    1390    &      0.0378    &   5      \\
59374.1558  & s        & 0.0007    &  701.5     &     0.0184     &    5    &   60108.3463  &   s      &  0.0003   &    1390.5  &      0.0394    &   5      \\
59374.6866  & p        & 0.0007    &  702       &     0.0165     &    5    &   60108.8779  &   p      &  0.0003   &    1391    &      0.0383    &   5      \\
59376.2871  & s        & 0.0007    &  703.5     &     0.0186     &    5    &   60112.6089  &   s      &  0.0003   &    1394.5  &      0.0398    &   5      \\
59376.8187  & p        & 0.0007    &  704       &     0.0174     &    5    &   60113.1408  &   p      &  0.0003   &    1395    &      0.0389    &   5      \\
59377.3523  & s        & 0.0007    &  704.5     &     0.0182     &    5    &   60113.6743  &   s      &  0.0003   &    1395.5  &      0.0397    &   5      \\
59377.8848  & p        & 0.0007    &  705       &     0.0180     &    5    &   60114.2067  &   p      &  0.0003   &    1396    &      0.0393    &   5      \\
59378.4176  & s        & 0.0007    &  705.5     &     0.0180     &    5    &   60114.7402  &   s      &  0.0003   &    1396.5  &      0.0400    &   5      \\
59378.9498  & p        & 0.0007    &  706       &     0.0175     &    5    &   60115.2715  &   p      &  0.0003   &    1397    &      0.0385    &   5      \\
59379.4835  & s        & 0.0007    &  706.5     &     0.0183     &    5    &   60115.8055  &   s      &  0.0003   &    1397.5  &      0.0397    &   5      \\
59380.0152  & p        & 0.0007    &  707       &     0.0173     &    5    &   60116.3371  &   p      &  0.0003   &    1398    &      0.0386    &   5      \\
59380.5492  & s        & 0.0007    &  707.5     &     0.0184     &    5    &   60116.8713  &   s      &  0.0003   &    1398.5  &      0.0400    &   5      \\
59381.0805  & p        & 0.0007    &  708       &     0.0170     &    5    &   60117.4024  &   p      &  0.0003   &    1399    &      0.0384    &   5      \\
59381.6147  & s        & 0.0007    &  708.5     &     0.0184     &    5    &   60117.9364  &   s      &  0.0003   &    1399.5  &      0.0395    &   5      \\
59382.1461  & p        & 0.0007    &  709       &     0.0170     &    5    &   60118.4677  &   p      &  0.0003   &    1400    &      0.0380    &   5      \\
59382.6802  & s        & 0.0008    &  709.5     &     0.0184     &    5    &   60119.5338  &   p      &  0.0003   &    1401    &      0.0386    &   5      \\
59383.2120  & p        & 0.0007    &  710       &     0.0173     &    5    &   60120.0683  &   s      &  0.0003   &    1401.5  &      0.0403    &   5      \\
59383.7465  & s        & 0.0007    &  710.5     &     0.0191     &    5    &   60120.5992  &   p      &  0.0003   &    1402    &      0.0384    &   5      \\
59384.2778  & p        & 0.0007    &  711       &     0.0176     &    5    &   60121.1337  &   s      &  0.0003   &    1402.5  &      0.0402    &   5      \\
59384.8103  & s        & 0.0007    &  711.5     &     0.0174     &    5    &   60121.6648  &   p      &  0.0003   &    1403    &      0.0385    &   5      \\
59385.3435  & p        & 0.0007    &  712       &     0.0178     &    5    &   60122.1997  &   s      &  0.0003   &    1403.5  &      0.0406    &   5      \\
59385.8765  & s        & 0.0007    &  712.5     &     0.0180     &    5    &   60122.7308  &   p      &  0.0003   &    1404    &      0.0389    &   5      \\
59386.4091  & p        & 0.0007    &  713       &     0.0178     &    5    &   60123.2646  &   s      &  0.0003   &    1404.5  &      0.0400    &   5      \\
59386.9421  & s        & 0.0007    &  713.5     &     0.0180     &    5    &               &          &           &            &                &          \\\hline
\end{longtable}
\textbf
{\footnotesize References:} \footnotesize (1) \citet{2009OEJV..116....1P}; (2) O - C gateway; (3) GCVS; (4) \citet{2010OEJV..130....1P};  (5) This work.

\begin{figure}
\centering
\includegraphics[width=0.8\linewidth]{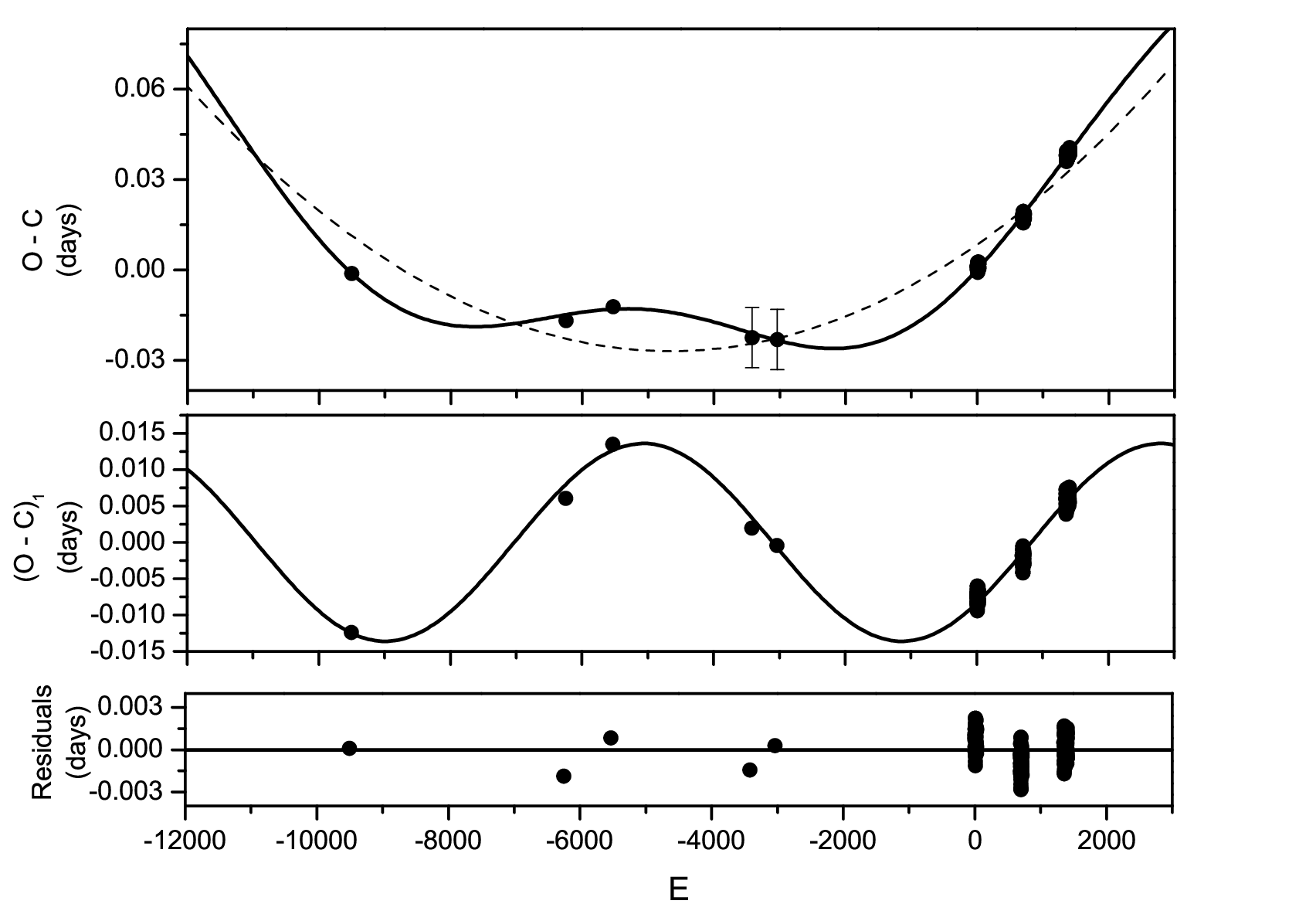}
\caption{The O - C diagram of NW Aps. In the upper panel, the solid line represents the theoretical O - C curve expressed by Equation \ref{New_ephemeris}, and the dashed line denotes to the quadratic part. In the middle panel, the periodic variation is clearly to be seen after the trend of quadratic variation is subtracted. The residuals are shown in the bottom after both quadratic and periodic variations are removed.}
\label{OC_fitting}
\end{figure}

The O - C diagram of NW Aps is displayed in the upper panel of Fig. \ref{OC_fitting}, and the solid circles refer to the $O - C$ values calculated from Equation \ref{Epoch_O-C} . The times of minima cover over the past 30 years. It is obviously to see that linear correction and parabola model can not explain the O - C variations well. Thus, a cyclic variation is also considered. The O - C variation can be expressed as follows:
\begin{equation}
O - C = \Delta T + \Delta P\times{E} + \frac{1}{2}\times{\frac{dp}{dt}}\times{E^2} + \tau,\label{Fitted-ephemeris}
\end{equation}
where $\Delta T$ and $\Delta P$ are corrections to the initial epoch and initial orbital period used in Equation \ref{Epoch_O-C}, and $\frac{dp}{dt}$ refers to the parabola changes in the O - C curve. $\tau$ represents the periodic variations resulting from light-travel time effect \citep{1952ApJ...116..211I} or the Applegate's mechanism \citep{1992ApJ...385..621A}, and detailed discussions can be found in Section 6.2. In our case, a circular orbit is reasonable to fit the O - C curve well. Therefore, eccentricity is assumed to be $e = 0$. The periodic variation $\tau$ in \citet{1952ApJ...116..211I} can be simplified as follows:
\begin{equation}
\tau = Ksin(\nu+\omega)
\end{equation}
and 
\begin{equation}
K = \frac{1}{2}[(O - C)_{max} - (O - C)_{min}] = \frac{a \cdot sini}{c},
\end{equation}
in which $K$ is the semi-amplitude of periodic variation and $c$ is the speed of light.
Finally, the new ephemeris is determined to be:
\begin{equation}\label{New_ephemeris}
\begin{array}{lll}
Min. I = 2458626.6569(\pm0.0001)+1.06557319(\pm0.00000006)\times{E} \\
         +1.630(\pm0.008)\times{10^{-9}}\times{E^{2}} \\
         + 0.0136(\pm0.0001)\times \sin[0.0458(\pm0.0002)\times E -37.8(\pm0.6)^{\circ} ]
\end{array}
\end{equation}

According to the newly determined ephemeris, the period of NW Aps is revealed to be increasing continuously at a rate of $dP/dt=+1.117(\pm0.005)\times{10^{-6}}day\cdot year^{-1}$. In the upper panel of Fig. \ref{OC_fitting}, the solid line represents the theoretical O - C curve expressed by Equation \ref{New_ephemeris}, and the dashed line denotes to the long-term increase in the orbital period. In the middle panel of Fig. \ref{OC_fitting}, the periodic variation is clearly to be seen after the trend of long-term increase is subtracted. The period of cyclic variation is calculated to be $P_3 = 22.9(\pm0.1)$ and the semi-amplitude is $K = 0.0136(\pm0.0001)$ day.

\section{Modeling of light curves}
\label{sect:lightcurve}
\begin{figure}
\centering
\includegraphics[width=0.8\linewidth]{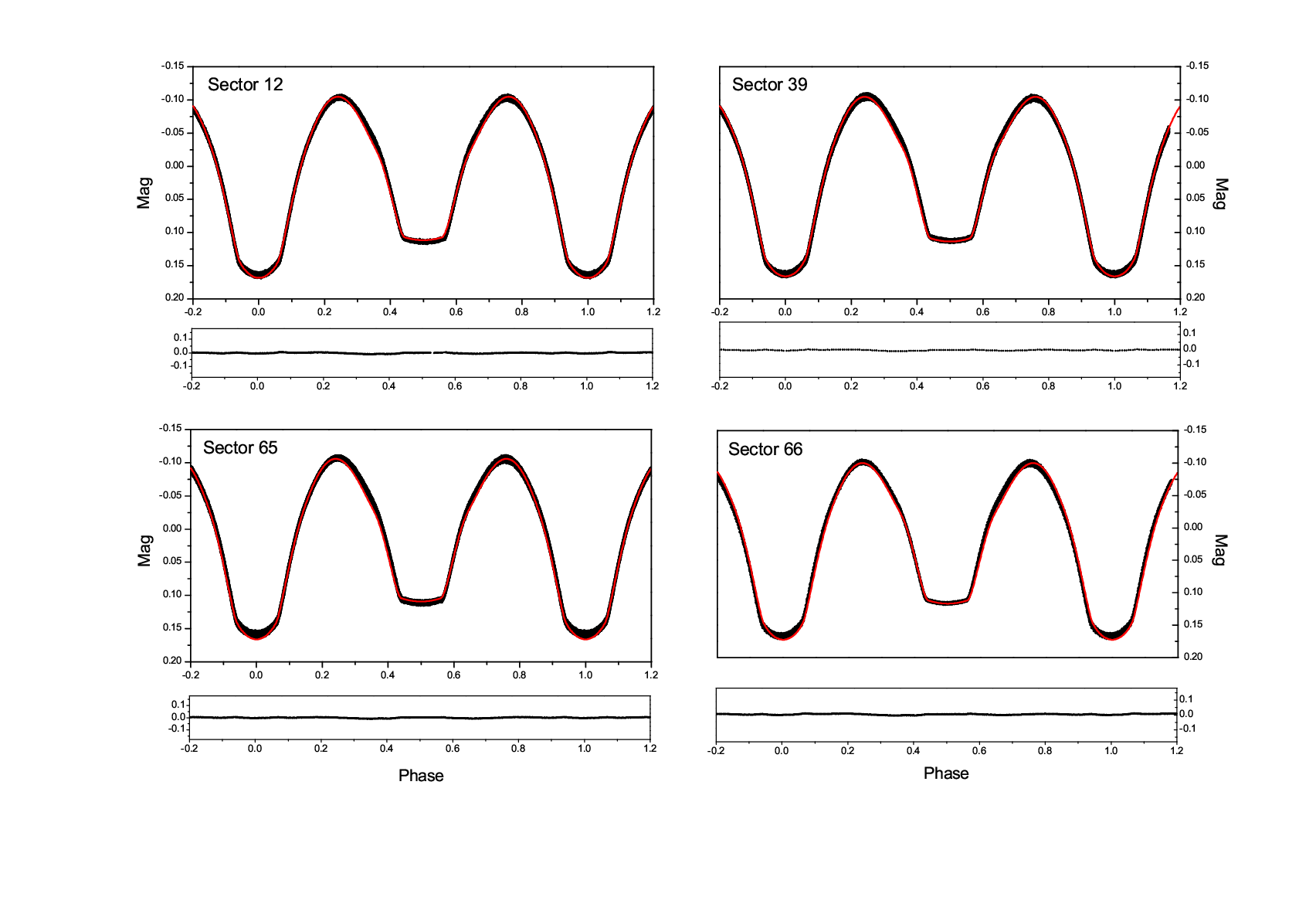}
\caption{The observational (black dots) and synthetic (red lines) light curves of NW Aps. The residuals are displayed under the light curves.}
\label{lc-fitting}
\end{figure}

The light curves displayed in Fig. \ref{lc} cover many orbital period circles. For each sector, we folded the observational data into one orbital circle. The PDCSAP flux is normalized and converted to magnitude as shown in Fig. \ref{lc-fitting}, and the solid circles represent TESS observational light curves. The Wilson-Devinney (WD) code is used to model the light curves \citep{2020ascl.soft04004W}. Mode 3 for contact configuration is applied since the light curves are typical EW-type. The mass ratio can be achieved with very high reliability since the light curves show totally eclipsing character \citep{2005Ap&SS.296..221T}. The effective temperature of the star eclipsed at the primary minimum is fixed at $ T_{1} = 5792K$, which was given in Gaia DR3 \citep{2023A&A...674A...1G}. The gravity-darkening coefficients are set at $g_{1} = g_{2} = 0.32$ \citep{1967ZA.....65...89L}. The bolometric albedo coefficients are set at $A_{1} = A_{2} = 0.5$ \citep{1969AcA....19..245R}. The limb darkening coefficients are set accordingly \citep{1993AJ....106.2096V}.  The adjustable parameters are mass ratio ($q$), orbital inclination ($i$), modified dimensionless surface potential of the primary star ($\Omega_{1}$), mean surface temperature of the secondary star ($T_{2}$), bandpass luminosity of the primary star ($L_{1}$). The third light ($l_{3}$) is also set to be a free parameter while running the code. However, the fitting results did not improve significantly. Considering that the parameter values of CROWDSAP in TESS Target Pixel Files' headers are over $97\,\%$ for NW Aps, the model without third light is adopted. The determined photometric parameters are listed in Table \ref{WD_results}. $\Omega_{2}$ refers to the modified dimensionless surface potential of the secondary star, and $\Omega_{1}=\Omega_{2}$ for contact model. $\Omega_{in}$ and $\Omega_{out}$ are the potential at the inner Lagrangian point and outer Lagrangian point, respectively. $\Delta T$ is the temperature difference between the primary and secondary star. $L_{2}$ represent the luminosity of the secondary star. $r_{1}(pole)$, $r_{1}(sde)$ and $r_{1}(back)$ are three characteristic radii of the primary star in Roche geometry, since stars in close binaries are non-spherical shape. $r_{2}(pole)$, $r_{2}(sde)$ and $r_{2}(back)$ are three characteristic radii of the secondary star. And all radii listed in Table \ref{WD_results} are scaled to the semimajor axis ($a$). The fill-out factor ($f$) is calculated with the equation $f = \frac{\Omega_{in} - \Omega_{1}}{\Omega_{in} - \Omega_{out}}$. The fitting residuals are listed in the last line. The theoretical light curves are displayed in Fig. \ref{lc-fitting} in red line.

\begin{table}
\begin{center}
\caption{Photometric solutions for NW Aps}\label{WD_results}
\setlength{\tabcolsep}{4pt}
\begin{tabular}{lllllllll}
\hline\hline
Parameters                 &   Sector 12              &   Sector 39                &   Sector 65             &   Sector 66               &    Averaged        \\\hline
$T_{1}(K)   $              &  5792(fixed)             &  5792(fixed)               &  5792(fixed)            &  5792(fixed)              &  5792(fixed)      \\
q ($M_2/M_1$ )             &  0.086($\pm0.001$)       &  0.086($\pm0.001$)         &  0.085($\pm0.001$)      &  0.087($\pm0.001$)        &  0.086($\pm0.001$)         \\
$i(^{\circ})$              &  80.6($\pm0.2$)          &  80.8($\pm0.6$)            &  80.8($\pm0.2$)         &  80.9($\pm0.3$)           &  80.8($\pm0.3$)      \\
$\Omega_{in}$              &  1.915                   &  1.915                     &  1.910                  &  1.917                    &  1.914      \\
$\Omega_{out}$             &  1.858                   &  1.859                     &  1.855                  &  1.860                    &  1.856     \\
$\Omega_{1}=\Omega_{2}$    &  1.870($\pm0.001$)       &  1.872($\pm0.005$)         &  1.865($\pm0.001$)      &  1.874($\pm0.002$)        &  1.870($\pm0.002$)    \\
$T_{2}(K)$                 &  5514($\pm6$)            &  5555($\pm13$)             &  5512($\pm4$)           &  5522($\pm6$)             &  5526($\pm7$)       \\
$\Delta T(K)$              &  278                     &  237                       &  280                    &  270                      &  266       \\
$T_{2}/T_{1}$              &  0.952($\pm0.001$)       &  0.959($\pm0.002$)         &  0.952($\pm0.001$)      &  0.953($\pm0.001$)        &  0.954($\pm0.001$)     \\
$L_{1}/(L_{1}+L_{2}$)      &  0.9051($\pm0.0001$)     &  0.9031($\pm0.0001$)       &  0.9056($\pm0.0001$)    &  0.9043($\pm0.0001$)      &  0.9045($\pm0.0001$)   \\
$l_{3}$                    &  0                       &  0                         &  0                      &  0                        &  0       \\
$r_{1}(pole)$              &  0.5572($\pm0.0005$)     &  0.5563($\pm0.0012$)       &  0.5583($\pm0.0001$)    &  0.5564($\pm0.0006$)      &  0.5571($\pm0.0006$)     \\
$r_{1}(side)$              &  0.6332($\pm0.0009$)     &  0.6316($\pm0.0021$)       &  0.6349($\pm0.0002$)    &  0.6318($\pm0.0010$)      &  0.6329($\pm0.0011$)     \\
$r_{1}(back)$              &  0.6538($\pm0.0009$)     &  0.6521($\pm0.0021$)       &  0.6556($\pm0.0002$)    &  0.6524($\pm0.0010$)      &  0.6535($\pm0.0011$)     \\
$r_{2}(pole)$              &  0.1983($\pm0.0035$)     &  0.1985($\pm0.0074$)       &  0.1985($\pm0.0002$)    &  0.1980($\pm0.0037$)      &  0.1983($\pm0.0037$)     \\
$r_{2}(side)$              &  0.2087($\pm0.0044$)     &  0.2090($\pm0.0092$)       &  0.2092($\pm0.0002$)    &  0.2084($\pm0.0046$)      &  0.2088($\pm0.0046$)     \\
$r_{2}(back)$              &  0.2724($\pm0.0189$)     &  0.2726($\pm0.0397$)       &  0.2772($\pm0.0009$)    &  0.2696($\pm0.0186$)      &  0.2730($\pm0.0195$)     \\
$f$                        &  $78.3\,\%(\pm4.0\,\%)$  &  $77.0\,\%(\pm8.8\,\%)$    &  $82.4\,\%(\pm0.7\,\%)$ &  $75.9\,\%(\pm4.3\,\%)$   &  $78.4\,\%(\pm4.5\,\%)$   \\
$\Sigma{\omega(O-C)^2}$    &  0.001183                &  0.001087                  &  0.001288               &  0.001229                 &  0.001197            \\
\hline
\hline
\end{tabular}
\end{center}
\end{table}

\section{Synthesis of spectral energy distribution}
\label{sect:sed}

\begin{figure}
\centering
\includegraphics[width=12cm, angle=0]{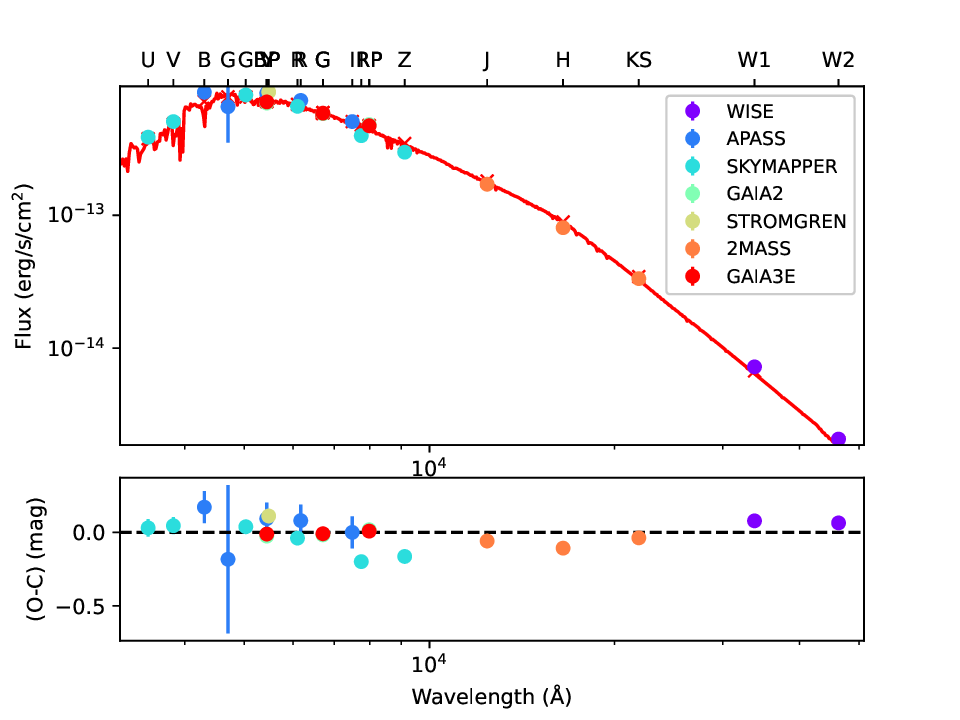}
\caption{SED fitting of NW Aps. The solid circles represent photometric observations and red line refers to theoretical spectral energy distributions (upper panel). The residuals are displayed in the lower panel.}
\label{fig:sed}
\end{figure}

\begin{figure}
\centering
\includegraphics[width=12cm, angle=0]{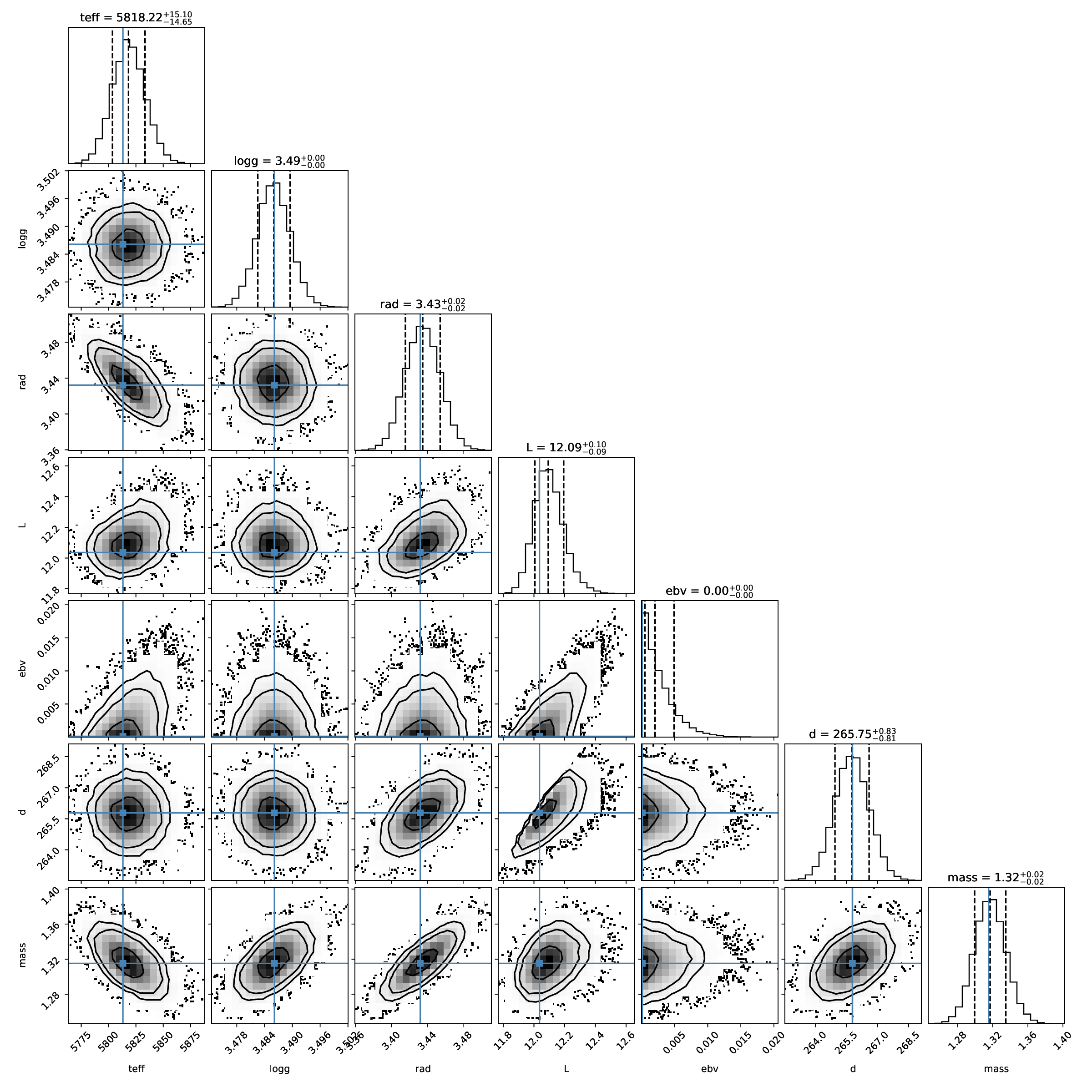}
\caption{Parameter distributions of the primary star in NW Aps derived from SED fitting.}
\label{fig:distribution}
\end{figure}

Spectral Energy Distribution (SED) shows the energy emitted by stars as a function of wavelength across the electromagnetic spectrum. It's a powerful tool to study the physical properties of stars. To fitting the SED of NW Aps, photometric observations from a variety of sky survey catalogs are combined, including SkyMapper \citep{2018PASA...35...10W}, Gaia Early Data Release 3 \citep{2021A&A...649A...1G}, Gaia Data Release 2 \citep{2018A&A...616A...1G}, AAVSO Photometric All-Sky Survey \citep{2009AAS...21440702H}, Str{\"o}mgren-Crawford uvby{\ensuremath{\beta}} photometry \citep{2015A&A...580A..23P}, Wide-field Infrared Survey Explorer \citep{2010AJ....140.1868W} and Two Micron All Sky Survey \citep{2006AJ....131.1163S}. Then, the Python package Speedyfit \footnote{https://speedyfit.readthedocs.io/en/latest/} is applied to fit the observations \citep{2012A&A...548A...6V}. As shown in Table \ref{WD_results}, the mass ratio of NW Aps is quite small and the secondary star contribute less than $10\,\%$ of the total luminosity. NW Aps shows a contact configuration, and both of the two component are embedded in their common envelope. Therefore, it is reasonable to assume that all the electromagnetic radiation come from the primary star and take the binary system as a single star during the fitting. The parallax, effective temperature and surface gravity are adopted from Gaia Data Release 3 \citep{2023A&A...674A...1G}, which are $p = 3.7507\pm0.0116 mas$, $T_{eff} = 5792^{-20}_{+23}K$ and $log g = 3.4858^{-0.0034}_{+0.0037}$. Then, the zero-point correction to the Gaia DR3 parallax is calculated to be $\Delta p = -0.0123 mas$ with the python package $gaiadr3\_zeropoint$\footnote{https://pypi.org/project/gaiadr3-zeropoint/}, which used the method described in \citet{2020A&A...633A...1L} \& \citet{2021A&A...649A...4L}. Thus, the parallax applied in the SED fitting is updated to be $p' = 3.7630\pm0.0116 mas$. The Markov chain Monte Carlo (MCMC) approach is used to fit the requested parameters and their errors. The optimal solution is displayed in Figure \ref{fig:sed}. The parameter distributions from MCMC approach is shown in Figure \ref{fig:distribution}. The mass of primary star is determined to be $M_1 = 1.32\pm0.02 M_\odot$. 

\section{Discussions}
\label{sect:discussions}

\subsection{Fundamental properties of NW Aps}
\label{subsect:fundermantal properties}
NW Aps is a quite bright contact binary in the southern celestial hemisphere and TESS obtained its lightcurves with very high precision. Considering its totally eclipsing character, we can conclude that the parameters determined in Table \ref{WD_results} are very reliable \citep{2021AJ....162...13L}. NW Aps is a low mass ratio ($q = 0.086$) contact binary system which has approached the minimum mass ratio of contact binary systems. The more fascinating fact is that it may be a stellar merger candidate with the longest orbital period to be reported so far. The mass of the more massive component in NW Aps is determined to be $M_{1} = 1.32(\pm0.02)M_\odot$ from SED fitting. The other absolute parameters of two component stars are calculated and listed in Table \ref{absolute}. The orbital semi-major axis is calculated to be $a = 4.95(\pm0.03)R_\odot$. 

The orbital period of NW Aps is increasing continuously at a rate of $dP/dt=+1.117(\pm0.005)\times{10^{-6}}day\cdot year^{-1}$. Considering the fact that mass transfer between the two component stars generally occurs through inner Lagrangian point ($L_2$) in contact and semi-contact binary systems, the long term period increasing is supposed to be caused by mass transfer from the less massive component to the more massive one. Assuming a conservative mass transfer case, the mass transfer rate is determined to be $\frac{dM_{2}}{dt}=-3.36(\pm0.02)\times{10^{-8}}M_\odot/year$ from the less massive secondary star to the more massive one according to Equation \ref{Masstransfer}. Rapid orbital period changes are expected before stellar merger events, such as in the case of V1309 Sco \citep{2011A&A...528A.114T} and KIC 9832227 \citep{2017ApJ...840....1M}, although the 2022 red nova merger prediction of KIC 9832227 was negated later \citep{2018ApJ...864L..32S}. Comparing the orbital period changes' rate of NW Aps with other W UMa-type contact binaries reported before \citep{2021ApJS..254...10L}, we can conclude that their period changes' rate are in the same time scale and not as rapid as expected. It seems that these low mass ratio contact binary systems are in a quite stable dynamical state. 

\begin{equation}\label{Masstransfer}
\frac{dM_{2}}{dt}=\frac{M_1M_2}{3p(M_2-M_1)}\times{dp/dt}
\end{equation}

\begin{table}
\caption{Absolute parameters of components in NW Aps}\label{absolute}
\begin{center}
\small
\begin{tabular}{lllllllll}
\hline
Parameters                        &Primary                         & Secondary          \\
\hline
$M$                               & $1.32(\pm0.02)M_\odot$         & $0.11\pm0.01)M_\odot$         \\
$R$                               & $3.04(\pm0.02)R_\odot$         & $1.11(\pm0.01)R_\odot$         \\
$L$                               & $9.35(\pm0.18)L_\odot$         & $1.04(\pm0.03)L_\odot$         \\
\hline
\end{tabular}
\end{center}
\end{table}

\subsection{Cyclic variation in the orbital period}
\label{subsect:cyclic}

The O - C curve in Figure \ref{OC_fitting} also reveals that a periodic variation of $P_3 = 22.9(\pm0.1)$ years is superposed on the quadratic changes. It may be resulted from the light-travel time effect (LTTE) of an additional stellar object orbiting around NW Aps. The third body is in a circular orbit, and the projected orbital radius that the central binary NW Aps orbit the barycenter of the triple system is  calculated to be
$a_{12} \cdot sini^{'} = K \times c$, in which $K$ is the semi-amplitude of periodic variation in Equation \ref{New_ephemeris}, $i^{'}$ is the orbital inclination of the potential third object, and $c$ is the speed of light. Thus, the projected orbital radius is calculated to be $a_{12}\cdot sini^{'} = 2.355(\pm0.017)AU$. Basing on the Kepler's third law, the mass function of the potential third object is
\begin{equation}\label{Massfunction}
f(m) = \frac{4\pi^2}{GP^2_3} \cdot (a_{12} \cdot sini^{'})^3 =\frac{(M_3 \cdot sini^{'})^3}{(M_1+M_2+M_3)^2}.
\end{equation}
$G$ is the gravitational constant, and the value is obtained to be $f(m) = 0.0249(\pm0.0006)$. Then, the lowest mass of the potential third object is calculated to be $M_3 = 0.436(\pm0.007)M_\odot$, and the orbital radius of the third body is less than $7.73(\pm0.11)AU$. Since there is no third light detected in the light curves, the potential third object should be a compact object, a white dwarf or a neutron star.

\begin{figure}
\centering
\includegraphics[width=0.8\linewidth]{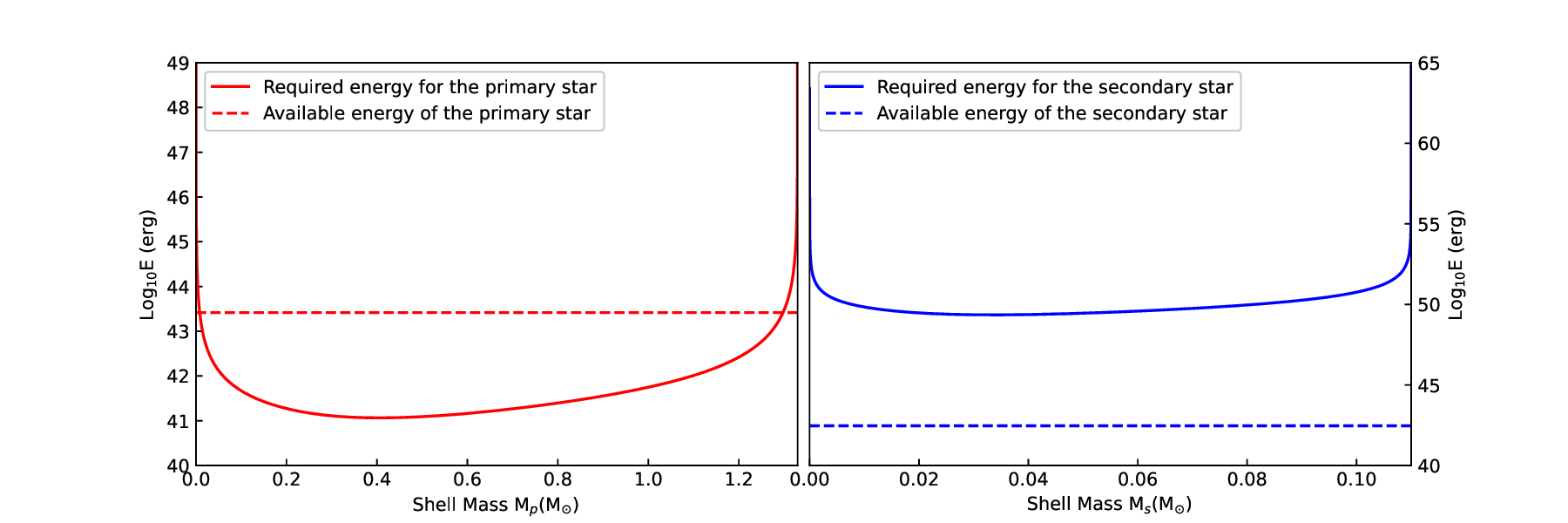}
\caption{The solid lines correspond to the energy required to generate the detected periodic variation applying the Applegate's mechanism. The dashed lines represent the radiant energy in a full magnetic activity cycle. Red and blue refer to primary star and secondary one in NW Aps, respectively.} 
\label{applegate}
\end{figure}

Both of the two component stars in NW Aps are solar-type stars, thus the Applegate's mechanism resulting from magnetic activity cycle may also cause cyclic variations in the orbital period \citep{1992ApJ...385..621A}. The Applegate's mechanism is a theoretical model that explains cyclic orbital period variations observed in close binary systems. Variations in the star's internal differential rotation and magnetic fields cause changes of its quadrupole moment, which leads to observable oscillations in the orbital period. The approach from \citet{2006MNRAS.365..287B} and \citet{2016A&A...587A..34V} are used to estimated the required energy for the cyclic change. As shown in Figure \ref{applegate}, the energy required to generate the detected periodic variation applying the Applegate's mechanism are figure out in solid lines and the radiant energy in a full magnetic activity cycle are plotted with dashed lines. It is obviously to state that the cyclic variation may also be cause by the magnetic activity of the primary star.

\subsection{Statistics research of extremely low mass ratio contact binaries}
\label{subsect:mass}

\begin{figure}
\centering
\includegraphics[width=0.8\linewidth]{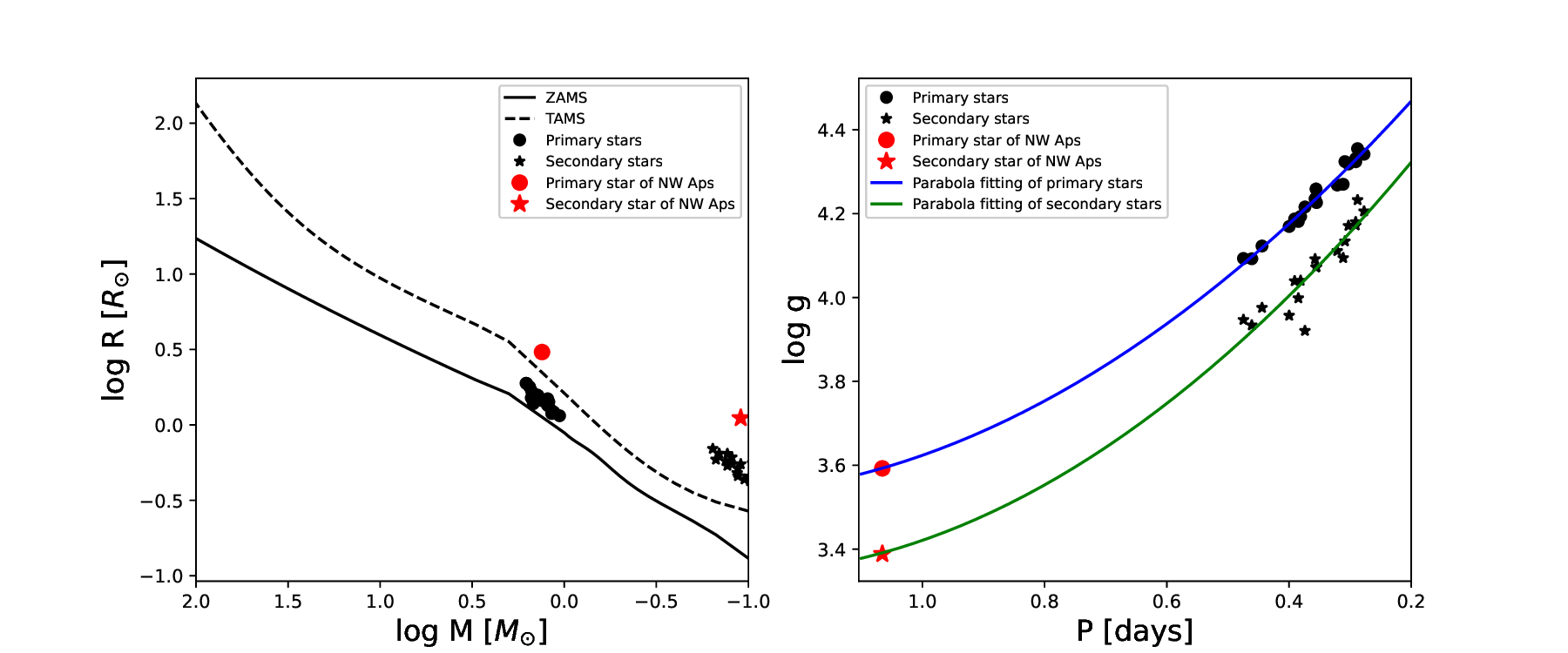}
\caption{The M - R and P - log g diagrams for totally eclipsing contact binaries with mass ratio $q < 0.1$.} 
\label{MR}
\end{figure}

Theoretical research predicts that low mass ratio contact binaries stay in the final evolutionary stage of short period contact binaries. They will merge into a rapidly rotating single star and form a FK Com-type star or a blue straggler in the center at last. All well studied totally eclipsing contact binaries with mass ratio $q < 0.1$ are collected and listed in Table \ref{lowmass}. They are stellar merger candidates and declared to be merging in the future except the case of V1309 Sco which has merged in 2008. It is easily to be seen that NW Aps has the longest orbital period among the stellar merger candidates and the orbital period of other candidates are shorter than 0.5 days. Although only one target is confirmed to date, the formation and evolution of long period low mass ratio contact binary is really intriguing to us. The surface gravity is calculated with the formula $g= \frac{GM}{R^2}$. The M - R and P - log $g$ relationships of targets listed in Table \ref{lowmass} (except V1309 Sco) are displayed in Fig. \ref{MR}. In the left panel, the solid line represents the Zero Age Main Sequence (ZAMS) line, while the dashed one refers to the Terminal Age Main Sequence (TAMS) line. The primary stars are still main sequence stars and the secondary stars is oversized to main sequence stars for low mass ratio contact binary with orbital period less than 0.5 days. The secondary stars may have evolved away from main sequence. As for NW Aps, both of the primary and secondary star are oversized to main sequence stars. In the right panel, the surface gravity of the primary and secondary stars in NW Aps are significant lower than main sequence stars. Both of the two component stars may have evolve into red giant stars. And it is obviously to see that there is a gap between short period low mass ratio contact binaries and NW Aps. The P - log g relation is fitted with parabola, which are displayed with blue and green lines for primary stars and secondary stars respectively. The fitting results are :
\begin{equation}\label{parabola-fitting-P}
\begin{array}{lll}
log g = 0.68(\pm0.09)\times{P^{2}}-1.87(\pm0.12)\times{P}+4.81(\pm0.03)
\end{array}
\end{equation}
for primary stars, and 
\begin{equation}\label{parabola-fitting-S}
\begin{array}{lll}
log g = 0.78(\pm0.25)\times{P^{2}}-2.06(\pm0.34)\times{P}+4.70(\pm0.09)
\end{array}
\end{equation}
for secondary stars.

\begin{table}
\caption{Totally eclipsing contact binaries with mass ratio $q < 0.1$.}\label{lowmass}
\begin{center}
\scriptsize
\begin{tabular}{lllllllllllllll}
\hline
Star                               & Period    &   q    &  $T_{1}$ &  $T_{2}$ &  f   &   $M_{1}$  &   $M_{2}$ &   $R_{1}$  &   $R_{2}$ &   $L_{1}$ &   $L_{2}$ &    $A$    & $dP/dt$    & Ref. \\
                                   & (days)    &        &    $K$   &   $K$    & $\%$ &  $M_\odot$ & $M_\odot$ & $R_\odot$  & $R_\odot$ & $L_\odot$ & $L_\odot$ & $R_\odot$ & $day\cdot year^{-1}$& \\\hline
V1309 Sco                          & 1.445600  &  0.094 &   4500   &  4354    &  89  &            &           &           &            &           &           &           &                        &  1 \\
NW Aps                             & 1.065558  &  0.086 &   5792   &  5526    &  60  &  1.32      &  0.11     &  3.04     &  1.11      &  9.35     &  1.04     &  4.95     & $+1.117\times{10^{-6}}$&  2 \\
CRTS J162327.1+031900              & 0.4745615 &  0.097 &   6914   &  6380    &  27  &  1.612     &  0.156    &  1.888    &  0.695     &  6.337    &  0.873    &  3.097    &                        &  3 \\
CRTS J155009.2+493639              & 0.4609100 &  0.082 &   6619   &  6199    &  19  &  1.599     &  0.130    &  1.882    &  0.644     &  6.429    &  0.773    &  3.015    & $+6.22\times{10^{-7}}$ &  3 \\
CRTS J223837.9+321932              & 0.4441697 &  0.093 &   6919   &  6756    &  45  &  1.541     &  0.144    &  1.784    &  0.646     &  5.463    &  0.704    &  2.916    & $+5.27\times{10^{-7}}$ &  3 \\
V870 Ara                           & 0.399722  &  0.082 &   5860   &  6210    &  96  &  1.503     &  0.123    &  1.67     &  0.61      &  2.96     &  0.50     &  2.686    &                        &  4 \\
CRTS J164000.2+491335              & 0.3907817 &  0.095 &   7137   &  6574    &  29  &  1.402     &  0.133    &  1.580    &  0.577     &  3.912    &  0.512    &  2.596    & $-3.19\times{10^{-7}}$ &  3 \\
1SWASP J132829.37+555246.1         & 0.384705  &  0.089 &   6300   &  6319    &  70  &  1.23      &  0.11     &  1.49     &  0.55      &  3.15     &  0.43     &  2.45     & $-4.46\times{10^{-7}}$ &  5 \\
V857  Her                          & 0.382230  &  0.065 &   8300   &  8513    &  84  &            &           &           &            &           &           &           & $+2.90\times{10^{-7}}$ &  6 \\
CRTS J155106.5+303534              & 0.3809891 &  0.089 &   7147   &  6819    &  69  &  1.384     &  0.122    &  1.559    &  0.552     &  3.806    &  0.468    &  2.537    & $-5.13\times{10^{-7}}$ &  3 \\
ZZ PsA                             & 0.37388   &  0.078 &   6514   &  6703    &  97  &  1.213     &  0.095    &  1.422    &  0.559     &  2.20     &  0.63     &  2.387    &                        &  7 \\
CRTS J160755.2+332342              & 0.3572878 &  0.099 &   7987   &  7476    &  36  &  1.310     &  0.130    &  1.445    &  0.537     &  3.053    &  0.414    &  2.394    &                        &  3 \\
ASASSN-V J071924.64+415705.4       & 0.355763  &  0.099 &   5894   &  6363    &  88  &  1.51      &  0.15     &  1.51     &  0.59      &  2.47     &  0.41     &  3.097    & $-6.70\times{10^{-7}}$ &  8 \\
CRTS J154254.0+324652              & 0.3549879 &  0.087 &   5885   &  6067    &  94  &  1.317     &  0.114    &  1.464    &  0.514     &  3.215    &  0.389    &  2.379    & $-8.44\times{10^{-7}}$ &  3 \\
CRTS J224827.6+341351              & 0.3210262 &  0.079 &   6077   &  5541    &  92  &  1.233     &  0.098    &  1.350    &  0.456     &  2.613    &  0.292    &  2.171    & $+7.71\times{10^{-6}}$ &  3 \\
ASAS J083241+2332.4                & 0.311321  &  0.065 &   6300   &  6667    &  51  &  1.22      &  0.08     &  1.34     &  0.42      &  2.66     &  0.25     &  2.111    & $+8.80\times{10^{-7}}$ &  9 \\
V1187 Her                          & 0.310766  &  0.044 &   6250   &  6651    &  80  &            &           &           &            &           &           &           & $-1.50\times{10^{-7}}$ &  10\\
V53 in NGC 6121                    & 0.308449  &  0.078 &   7415   &  6611    &  69  &  1.472     &  0.115    &  1.383    &  0.481     &  7.306    &  0.465    &  2.24     & $-5.89\times{10^{-7}}$ &  11\\
CRTS J133031.1+161202              & 0.3026661 &  0.098 &   5860   &  6019    &  56  &  1.162     &  0.114    &  1.237    &  0.459     &  2.018    &  0.271    &  2.059    &                        &  3 \\
CRTS J170307.9+020101              & 0.2908836 &  0.093 &   5433   &  5237    &  70  &  1.134     &  0.105    &  1.204    &  0.436     &  1.874    &  0.271    &  1.985    & $+2.54\times{10^{-7}}$ &  3 \\
CRTS J234634.7+222824              & 0.2906934 &  0.086 &   5851   &  5746    &  37  &  1.141     &  0.098    &  1.218    &  0.425     &  1.953    &  0.233    &  1.964    & $+2.19\times{10^{-7}}$ &  3 \\
NSVS 2569022                       & 0.287797  &  0.078 &   6100   &  6100    &   1  &  1.17      &  0.09     &  1.19     &  0.38      &  1.73     &  0.17     &  1.98     &                        &  12\\
VSX J082700.8+462850               & 0.277158  &  0.055 &   5870   &  5728    &  19  &  1.06      &  0.06     &  1.15     &  0.32      &  1.40     &  0.11     &  1.83     & $-9.52\times{10^{-7}}$ &  5 \\
\hline
\end{tabular}
\end{center}
\textbf
{\footnotesize References:} \footnotesize (1) \citet{2016RAA....16...68Z}; (2) the present work; (3) \citet{2023MNRAS.519.5760L}; (4) \citet{2007A&A...465..943S}; (5) \citet{2021ApJ...922..122L};  (6) \citet{2005AJ....130.1206Q}; (7) \citet{2021MNRAS.501..229W}; (8) \citet{2022AJ....164..202L}; (9) \citet{2016AJ....151...69S}; (10) \citet{2019PASP..131e4203C}; (11) \citet{2017PASJ...69...79L}; (12) \citet{2018RAA....18..129K}.
\end{table}

The ratio of the spin angular momentum to the orbital angular momentum can be calculated with Equation \ref{angular} from \citet{1998CoSka..28..101P}, in which $r_{1,2}$ and $k_{1,2}$ are the relative radii and dimensionless gyration radii for both component stars respectively. The values are calculated to be $\frac{J_{spin}}{J_{orb}}$ = 0.307  or $\frac{J_{orb}}{J_{spin}}$ = 3.257 while $k_{1}^{2}$ = $k_{2}^{2}$ = 0.06 are assumed \citep{2006MNRAS.369.2001L}. Thus, NW Aps is still in a stable state since its orbital angular momentum ($J_{orb}$) is larger than 3 times of the total spin angular momentum ($J_{spin}$). NW Aps may become unstable and merge as in the case of V1309 Sco at last, which are predicted by the evolutionary theory of low mass ratio contact binaries. However, considering the particularity in NW Aps, its real evolutionary scenario is still uncertain to us. More photometric and spectroscopic observations will be carried out to reveal some more details on this target in the future. And also, more long period low mass ratio contact binaries, particularly targets in the gap, are needed to be confirmed from sky survey data, which may have great significance in understanding the formation and evolution of long period low mass ratio contact binary.

\begin{equation}\label{angular}
\frac{J_{spin}}{J_{orb}}=\frac{1+q}{q}\times{[(k_{1}r_{1})^{2} + q(k_{2}r_{2})^{2}]}
\end{equation}

\section{Conclusions}
\label{sect:conclusions}

Basing on the solutions derived from TESS light curve and SED fitting, we conclude that NW Aps is an extremely low mass ratio solar-type contact binary. It may lie in the final evolutionary stage of low mass binary systems, and may merger into a rapidly rotating star accompany with luminous optical outburst as theoretical predicted \citep{1980A&A....92..167H}. Stellar mergers are extremely energetic events in the Universe, normally accompanying with strong outburst and leading to the formation of peculiar stars, or sometimes triggering a supernova or other explosive phenomena \citep{2012ARA&A..50..107L,2020RAA....20..161H,2024ARA&A..62...21M}. Merger of compact binaries have gained plenty of attentions since they are typical potential gravitational wave candidates \citep{2016PhRvL.116x1102A,2017ApJ...848L..13A}. However, our knowledge in relation to merger of low mass binary system is still far from adequate. It's very hard to catch the merger events of low mass binary like V1309 Sco. Thus, efforts are trying to searching their potential progenitor, namely extremely low mass ratio contact binary systems \citep{2015AJ....150...69Y,2022MNRAS.512.1244C}. In the present work, we can see that almost all reported extremely low mass ratio contact binary are in a quite short orbital period less than 0.5 day. NW Aps is the only stellar merger candidate with orbital period longer than 1 day, and it indeed demonstrate some peculiarities, making it worth to study further. In the next step, we will make more efforts in capturing stellar merger candidates or even stellar merge events to enrich our understanding in low mass star evolutions.

\section*{Acknowledgements} 
I sincerely appreciate the reviewer’s thoughtful and valuable suggestions, which have significantly enhanced the quality of this work. This work is supported by the International Partnership Program of Chinese Academy of Sciences (No. 020GJHZ2023030GC), the Young Talent Project of Yunnan Revitalization Talent Support Program, and the Yunnan Fundamental Research Projects (Grant Nos. 202401AW070004, 202503AP140013, 202401AS070046). This research has made use of the SIMBAD database, operated at CDS, Strasbourg, France. This paper has made use of data collected by the Transiting Exoplanet Survey Satellite (TESS) mission, for which funding is provided by the NASA Explorer Program. All the {\it TESS} data used in this paper can be found in MAST: {10.17909/t9-nmc8-f686}.

\section*{Data Availability}
The TESS data are available at https://archive.stsci.edu/missions-and-data/tess. All data used in the present paper will be shared on reasonable request to the corresponding author.



\bibliographystyle{mnras}
\bibliography{references} 








\bsp	
\label{lastpage}
\end{document}